# A solid-fluid mixture model allowing for solid dilatation under external pressure

**Giulio Sciarra**[1], **Francesco dell'Isola**[1], **Kolumban Hutter**[2]

[1] Dipartimento di Ingegneria Strutturale e Geotecnica, Università di Roma *La Sapienza* Via Eudossiana 18, 00184 Roma, Italia
[2] Department of Mechanics, Darmstadt University of Technology, 64289 Darmstadt, Germany

A sponge subjected to an increase of the outside flui pressure expands its volume but nearly mantains its true density and thus gives way to an increase of the interstitial volume. This behaviour, not yet properly described by solid-flui mixture theories, is studied here by using the Principle of Virtual Power with the most simple dependence of the free energy as a function of the partial apparent densities of the solid and the fluid The model is capable of accounting for the above mentioned dilatational behaviour, but in order to isolate its essential features more clearly we compromise on the other aspects of deformation. Specifically the following questions are addressed: (i) The boundary pressure is divided between the solid and flui pressures with a dividing coefficien which depends on the constituent apparent densities regarded as state parameters. The work performed by these tractions should vanish in any cyclic process over this parameter space. This condition severely restricts the permissible constitutive relations for the dividing coefficient which results to be characterized by a single material parameter. (ii) A stability analysis is performed for homogeneous, pressurized reference states of the mixture by postulating a quadratic form for the free energy and using the afore mentioned permissible constitutive relations. It is shown that such reference states become always unstable if only the external pressure is sufficientl large, but the exact value depends on the interaction terms in the free energy. The larger this interaction is, the smaller will be the critical (smallest unstable) external pressure. (iii) It will be shown that within the stable regime of behaviour an increase of the external pressure will lead to a decrease of the solid density and correspondingly an increase of the specifi volume, thus proving the wanted dilatation effects. (iv) We close by presenting a formulation of mixture theory involving second gradients of the displacement as a further deformation measure (Germain 1973); this allows for the regularization of the otherwise singular boundary effects (dell'Isola and Hutter 1998, dell'Isola, Hutter and Guarascio 1999).

## 1 Introduction

In many engineering applications of binary mixture models of solid-flui interactions the pore space or the permeabilities are prescribed functions of the spatial coordinates, but they do not evolve together with the temporal changes of the other fiel variables: the engineering models based on this assumption stem from the studies of Terzaghi [23]. One example to the contrary is for instance the slow creeping deformation and the percolation of brine through a salt formation from a pressurized cavern fille with a liquid (for a detailed discussion of these phenomena see for instance [5]). Observations indicate that an increase in cavern pressure will not only result in a very slow creeping deformation of the salt dome but equally enhance the diffusion of liquid through the salt. This increase in percolation is not only due to an increase of the partial pressure of the liquid as a reaction to an increase of the cavern pressure but also because of grain boundary openings in the vicinity of the cavern walls.

Many of the presently existing mixture theories treating solid-flui interactions may in principle be able to cope with the flui dilatation mechanism: Bowen [1], [2], Müller [16], Morland [15], Rajagopal & Tao [17], Svensdsen & Hutter [22]; scrutiny has shown, however, that very particular constitutive behaviour must be assumed to achieve it. Terzaghi and Fillunger were aware that their models were unable to describe this dilatancy (see for instance de Boer & Ehlers [7]), Svendsen & Hutter's [22] mixture theory allows for it, but as shown by [10],[11] rather singular functional forms of the thermodynamic free energies $\psi(\cdot, n)$ as functions of the porosity $n$ are required if an appreciable space opening of the pores is to be achieved. It was then thought that introducing density gradients as independent constitutive variables would regularize the formulation [9], and indeed it did so.

However, in the problem stated above and in the mentioned papers we were confronted with a further difficulty the flu boundary conditions between the single constituent body and the mixture. As shown by Hutter et al. [14] the jump condition of momentum requires non-vanishing momentum production terms to exist on the singular surface (see also Svendsen & Gray [21]). An alternative, simpler than this, and not requiring surface balances is to postulate a parametrization how the traction on the single-constituent side of the surface is distributed between the tractions on the mixture-side [18]. This parametrization can be expressed as a scalar parameter (akin to surface fraction) depending on a number of variables, say the constituent densities. Obviously, the parametrization must be such that the work done by these boundary tractions in a simply connected closed circuit in this parameter space is zero for otherwise a perpetuum mobile of the firs kind would result (Seppecher, personal communication). The construction of the potential for the boundary tractions restricts the parametrization in our case to the extent that the functional dependence is fixe except for a single constant.

A mixture of two constituents one of which is a flui can only exist in equilibrium when it is confined i.e., when a pre-stress is exerted on it; this means that reference states with non-vanishing flui densities are always pre-stressed in such media. This fact gives rise to the question how the free energy describing the interior behaviour of the mixture must be structured that perturbations about such pre-stressed states are stable. For a quadratic dependence of the free energy upon the constituent densities the analysis shows that the stability condition depends strongly on the coupling term involving the two densities. For isotropic stresses and a homogeneous reference state we will prove that there is always an upper bound of the external pressure beyond which such states become unstable, corroborating Fillunger and Terzaghi's explosion under pressure (de Boer [8]).

One of the significan results determined in this paper is the fact that, depending on the coefficient of the parabolic representation of the free energy, a stable perturbation of a homogeneous pre-stressed reference state can give rise to a decrease in the apparent solid density with an increase of the external pressures. If the true density of the flui is essentially constant this corresponds to a dilatation of the space occupied by the fluid This is essentially how a sponge responds to the absorption of water from the outside; however, this property is exactly what is needed to achieve an increase of permeability without phase changes of the salt in the salt cavern problem mentioned at the beginning. The spatially dependent pore opening close to the cavern space is manifest as a boundary layer (in what is called Disturbed Rock Zone) and for this reason we close this paper with a presentation of the higher gradient model corresponding to the analysis of the earlier sections. A one-dimensional linear-elastic problem is finall presented as a firs application of the newly introduced model: we prove that close to unstable pre-stressed reference configuration the thickness of the boundary layer (where the apparent mass densities are not constant) at the external interface of the solid-flui mixture tends to infinity The regularizing properties of the proposed second gradient model are thus established.

The ultimate problem will have to incorporate visco-plastic components to also account for the creeping deformation of the salt (see Cosenza [5]).

## 2 Theoretical concepts

Consider a binary mixture of a solid matrix with connected pores which are fille  with a liquid. This arrangement can be thought of as a soil, rock or sponge. Let the two components be referred to as the solid and the flui  and indicate them by the suffice  $s$ and $f$. Let, moreover $\rho_s, \rho_f$ and $\mathbf{v}_s, \mathbf{v}_f$ be the solid and the flui  apparent densities and velocities, respectively, in the mixture. The mixture density and the barycentric velocity are then given by

$$\rho = \rho_s + \rho_f \Rightarrow 1 = \frac{\rho_s}{\rho} + \frac{\rho_f}{\rho} =: \xi_s + \xi_f, \tag{1}$$

$$\mathbf{v} = \xi_s \mathbf{v}_s + \xi_f \mathbf{v}_f, \tag{2}$$

in which $\xi_s$ and $\xi_f$ are the mass fractions of the solid and the fluid respectively.

We conceive this mixture to be non-reactive so that the balances of mass for the constituents reduce to

$$\frac{\partial \rho_a}{\partial t} + div(\rho_a \mathbf{v}_a) = 0, \quad (a = s, f). \tag{3}$$

In the ensuing analysis we shall restrict ourselves to purely mechanical processes; temperature will play no role, and so the constituent momentum equations are the only additional dynamical equations to be added to (3). Instead of a direct application of these laws we shall use the Principle of Virtual Power (PVP) applied to the appropriate energy functional to derive them. Let $\psi^{int}$ be the energy volume density and, as we limit ourselves to spherical states, assume it to be depending on the densities $\rho_a$: $\psi^{int} = \psi^{int}(\rho_s, \rho_f)$. Other dependencies could be incorporated, but will not be here for simplicity. The PVP states that the variation of the total energy in the considered body (a mixture) related to its (barycentric) motion equals the power of the external forces. If the exterior of the mixture body is a flui  then the boundary traction exerted on the mixture is a pressure, $p^{ext}$ which must be distributed between the constituents *via* the phenomenological *ansatz*

$$p_f = d_f \, p^{ext}, \qquad p_s = d_s \, p^{ext}, \qquad d_s + d_f = 1, \tag{4}$$

in which $d_s$ and $d_f$ are surface fraction parameters which, like the energy itself, depend upon the densities $\rho_s$ and $\rho_f$. This parametrization is of constitutive nature and has been proposed already in the past (Morland [15], Rajagopal & Tao [17]). Generally the areal fractions are identifie  with the volume fractions. This identificatio  traces back in mining engineering to the law of Delesse (1848), as quoted by de Boer [8]. In this paper we demonstrate the interrelation of the two via an argument concerning the firs  law of therodynamics. Thus the PVP yields the statement

$$\frac{d}{dt} \int_{\mathcal{B}} \psi^{int} dV = -\int_{\partial \mathcal{B}} d_a \, p^{ext} \, \mathbf{n} \cdot \mathbf{v}_a \, dA, \tag{5}$$

where summation over doubly repeated indices $a$ is understood and $\mathbf{n}$ denotes the outward pointing unit normal to $\partial \mathcal{B}$ which is the boundary of the Eulerean region $\mathcal{B}$, the actual placement of the mixture body.

We assume the external pressure $p^{ext}$ to be a conservative field  This means that the work performed by $p^{ext}$ on a cyclic quasi-static variation of the state parameters $\rho_s$ and $\rho_f$ must be path independent for otherwise a perpetuum mobile of the firs  kind would emerge. This is avoided if a potential $\psi^{ext}$ exists such that[1]

$$\frac{d}{dt} \int_{\mathcal{B}} \psi^{ext} dV = -\int_{\partial \mathcal{B}} d_a \, p^{ext} \, \mathbf{n} \cdot \mathbf{v}_a \, dA. \tag{6}$$

Employing the Reynolds Transport Theorem on the LHS, the divergence theorem on the RHS of (6) and using (3) yields

$$\int_{\partial \mathcal{B}} \left( -\rho_a \frac{\partial \psi^{ext}}{\partial \rho_a} + \xi_a \psi^{ext} + d_a p^{ext} \right) \mathbf{n} \cdot \mathbf{v}_a dA + \int_{\mathcal{B}} \rho_a \nabla \left( \frac{\partial \psi^{ext}}{\partial \rho_a} \right) \cdot \mathbf{v}_a \, dV = 0. \tag{7}$$

---

[1] We are grateful to P. Seppecher for drawing our attention to this fact.

Since (7) must hold for all velocity field $\mathbf{v}_a$, using a localization procedure the following restrictions on constitutive equations emerge:

$$\frac{\partial \psi^{ext}}{\partial \rho_a} = \frac{1}{\rho}\psi^{ext} + \frac{d_a}{\rho_a}p^{ext}, \tag{8}$$

$$\rho_a \nabla \left(\frac{\partial \psi^{ext}}{\partial \rho_a}\right) = 0 \quad \Longrightarrow \quad \frac{\partial \psi^{ext}}{\partial \rho_a} = p^{ext} c_a, \tag{9}$$

where $p^{ext} c_\alpha$ are constants and $p^{ext}$ is inserted for convenience. Each of these statements constitutes two equations and from their exploitation $\psi^{ext}$ and $d_a$ can be determined. They can be regarded as a vectorial statement in the 2D state-variable space spanned by $\rho_s$ and $\rho_f$. The second one implies that ($c_s, c_f$ and $c$ being integration constants)

$$\psi^{ext}(\rho_s, \rho_f) = p^{ext}\left(c_s \rho_s + c_f \rho_f + c\right), \tag{10}$$

so that using this in (8) we obtain

$$d_s = \xi_s + \frac{\rho_s}{[\rho]}(1 - \xi_s) = \xi_s\left(1 + \frac{\rho_f}{[\rho]}\right), \quad c = -1, \tag{11}$$

where

$$(c_s - c_f) =: \frac{1}{[\rho]}. \tag{12}$$

Let us pause to interpret this result: the requirement that the external pressure does not perform work along a closed trajectory in the state space $(\rho_s, \rho_f)$ has led to a restriction on the constitutive equations for $d_s$ (and $d_f$), which divide the external pressure $p^{ext}$ between the partial pressures $p_s$ and $p_f$, respectively. The division $p_a = \xi_a p^{ext}$ is an obvious and allowable possibility ($[\rho] \to \infty$), but it is not exhaustive. Any choice obeying (8) and (9) must have the form (11) in which the quantity $\rho_s/[\rho]$ is a scale parameter controlling how the boundary pressure is divided between the constituents.

Recall that $0 < d_s < 1$ and $0 < d_f < 1$ which imply

$$-\frac{1}{\rho_f} < \frac{1}{[\rho]} < \frac{1}{\rho_s}. \tag{13}$$

Evidently the typical scale parameter for the division of the pressure $p^{ext}$ between the constituent pressures involves an inverse density. Obviously the mixture density in the reference configuratio is a permissible choice for $[\rho]$, in principle it may be infinitel  large; $d_s$ then lies within the interval $0 < d_s < 1$ and is a linear function of $\xi_s$ between these two values.

Observe that the condition $d_a = 0$ implies that either $\rho_a = 0$ or $\rho - \rho_a = [\rho]$. The second of these conditions is physically impossible as both $\rho_a$ and $\rho$ must be considered as variables. This means that when one part of the external pressure $p^{ext}$ acting on the flui  (or on the solid) is equal to zero then the fluid-soli  mixture at the boundary reduces to the solid skeleton (or to a pure fluid  only.

## 3 Linearization in the neighbourhood of a pre-stressed reference state

Let $\mathscr{B}_0$ be the mixture body in its reference state with positions labeled by $\mathbf{X}$. Constituent densities and the external pressure in this state are denoted by $\rho_a^0$ ($a = s, f$) and $p_0^{ext}$, respectively; they will be assumed to be different from zero, and thus will give rise to constituent stress distributions $p_a^0$. Such a nontrivial pre-stressed situation is a necessary requirement for the mixture material introduced in Sect. 2. Indeed, the model is only meaningful when $\rho_s > 0$ and $\rho_f > 0$; thus $d_s > 0$ and $d_f > 0$, and this necessarily implies, through the parametrization of the constituent pressures that $p_s^0 \neq 0$ and $p_f^0 \neq 0$, if $p_0^{ext} \neq 0$. For the flui  to exist at all, a nonvanishing pressure is mandatory. So only nontrivial pre-stressed conditions are possible.

Consider a pre-stressed reference configuratio  of the solid and suppose the displacement of the material particles of the solid constituent to be small so that geometric linearizations are justifie  throughout. Reference values of fiel  quantities will carry the suffi  $(\cdot)_0$, perturbed field  will be denoted by a tilde, $(\widetilde{\cdot})$; therefore, in view of (10) we have

$$\begin{aligned}\psi^{ext} &= p_0^{ext}\left(c_s\rho_s^0 + c_f\rho_f^0 - 1\right) \\ &+ p_0^{ext}\left(c_s\tilde{\rho}_s + c_f\tilde{\rho}_f\right) + \tilde{p}^{ext}\left(c_s\rho_s^0 + c_f\rho_f^0 - 1\right) \\ &+ \tilde{p}^{ext}\left(c_s\tilde{\rho}_s + c_f\tilde{\rho}_f\right).\end{aligned} \quad (14)$$

Alternatively, we express the potential $\psi^{int}$ as a quadratic form about the reference state $\rho_s^0$, $\rho_f^0$; this will be done in the form $\psi^{int} = \rho\varphi^{int}$, in which $\varphi^{int}$ is the internal energy per unit mass. What obtains is as follows:

$$\begin{aligned}\psi^{int} &= \rho\varphi^{int}\left(\rho_s,\rho_f\right) \\ &\simeq (\rho_0 + \tilde{\rho})\left(\gamma_s\tilde{\rho}_s + \gamma_f\tilde{\rho}_f + \frac{1}{2}\gamma_{ss}\tilde{\rho}_s^2 + \gamma_{sf}\tilde{\rho}_s\tilde{\rho}_f + \frac{1}{2}\gamma_{ff}\tilde{\rho}_f^2 + ...\right) \\ &= \rho_0\left(\gamma_s\tilde{\rho}_s + \gamma_f\tilde{\rho}_f\right) + \frac{1}{2}(2\gamma_s + \rho_0\gamma_{ss})\tilde{\rho}_s^2 + \frac{1}{2}(2\gamma_f + \rho_0\gamma_{ff})\tilde{\rho}_f^2 \\ &+ (\gamma_s + \gamma_f + \rho_0\gamma_{sf})\tilde{\rho}_s\tilde{\rho}_f,\end{aligned} \quad (15)$$

where an arbitrary constant in $\varphi^{int}$ is irrelevant and where the constants

$$\begin{aligned}\gamma_s &:= \left.\frac{\partial\varphi^{int}}{\partial\rho_s}\right|_{\rho_s^0,\rho_f^0}, \quad \gamma_f := \left.\frac{\partial\varphi^{int}}{\partial\rho_f}\right|_{\rho_s^0,\rho_f^0}, \\ \gamma_{ss} &:= \left.\frac{\partial^2\varphi^{int}}{\partial\rho_s^2}\right|_{\rho_s^0,\rho_f^0}, \quad \gamma_{ff} := \left.\frac{\partial^2\varphi^{int}}{\partial\rho_f^2}\right|_{\rho_s^0,\rho_f^0}, \quad \gamma_{sf} := \left.\frac{\partial^2\varphi^{int}}{\partial\rho_s\partial\rho_f}\right|_{\rho_s^0,\rho_f^0}\end{aligned} \quad (16)$$

are supposed known when the reference state is known. Depending upon the value of $\gamma_{sf}$ relative to the values of $\gamma_{ss}$, $\gamma_{ff}$ coupling will be called weak or strong. Generally growing $\gamma_{sf}$ increases this coupling.

Next, we wish to write down the equilibrium equations which follow from the localization of

$$\frac{d}{dt}\int_{\mathcal{B}}\psi^{tot}dV = \frac{d}{dt}\int_{\mathcal{B}}\left(\psi^{int} - \psi^{ext}\right)dV = 0. \quad (17)$$

For static conditions this yields

$$\begin{aligned}\rho_s\frac{\partial\psi^{int}}{\partial\tilde{\rho}_s} - \xi_s\psi^{int} &= \rho_s\frac{\partial\psi^{ext}}{\partial\tilde{\rho}_s} - \xi_s\psi^{ext}, \\ \rho_f\frac{\partial\psi^{int}}{\partial\tilde{\rho}_f} - \xi_f\psi^{int} &= \rho_f\frac{\partial\psi^{ext}}{\partial\tilde{\rho}_f} - \xi_f\psi^{ext}.\end{aligned} \quad (18)$$

Using (14) and (15) and expanding as illustrated above yields the following zeroth and firs  order problems:

*zeroth order*

$$\gamma_s = \frac{p_0^{ext}}{\rho_0^2} + \frac{\rho_f^0}{\rho_0^2}(c_s - c_f)p_0^{ext}, \qquad \gamma_f = \frac{p_0^{ext}}{\rho_0^2} - \frac{\rho_s^0}{\rho_0^2}(c_s - c_f)p_0^{ext}. \quad (19)$$

*first order*

$$\begin{bmatrix} R_{11} & R_{12} \\ R_{21} & R_{22} \end{bmatrix}\begin{bmatrix} \tilde{\rho}_s \\ \tilde{\rho}_f \end{bmatrix} = \begin{bmatrix} \xi_s^0 + \dfrac{\rho_s^0\rho_f^0}{\rho_0}(c_s - c_f) \\ \xi_f^0 - \dfrac{\rho_s^0\rho_f^0}{\rho_0}(c_s - c_f) \end{bmatrix}\tilde{p}^{ext}, \quad (20)$$

where

$$R_{11} := \rho_s^0 \left(2\gamma_s + \rho_0\gamma_{ss}\right) + \rho_f^0\gamma_s - p_0^{ext}\left(\left(\xi_f^0\right)^2 (c_s - c_f) + \frac{\xi_f^0}{\rho_0}\right),$$

$$R_{12} := \rho_s^0 \left(\gamma_s + \gamma_f + \rho_0\gamma_{sf}\right) - \rho_s^0\gamma_f - p_0^{ext}\left(\left(\xi_s^0\right)^2 (c_s - c_f) - \frac{\xi_s^0}{\rho_0}\right),$$

$$R_{21} := \rho_f^0 \left(\gamma_s + \gamma_f + \rho_0\gamma_{sf}\right) - \rho_f^0\gamma_s + p_0^{ext}\left(\left(\xi_f^0\right)^2 (c_s - c_f) + \frac{\xi_f^0}{\rho_0}\right),$$

$$R_{22} := \rho_f^0 \left(2\gamma_f + \rho_0\gamma_{ff}\right) + \rho_s^0\gamma_f + p_0^{ext}\left(\left(\xi_s^0\right)^2 (c_s - c_f) - \frac{\xi_s^0}{\rho_0}\right). \quad (21)$$

Equations (19) relate $\gamma_s$ and $\gamma_f$ to one another. On the other hand, (20) could be inverted to obtain $\tilde{\rho}_s$ and $\tilde{\rho}_f$ in terms of $\tilde{p}^{ext}$

$$\begin{bmatrix} \tilde{\rho}_s \\ \tilde{\rho}_f \end{bmatrix} = [\mathbf{R}]^{-1} \begin{bmatrix} \xi_s^0 + \frac{\rho_s^0\rho_f^0}{\rho_0}(c_s - c_f) \\ \xi_f^0 - \frac{\rho_s^0\rho_f^0}{\rho_0}(c_s - c_f) \end{bmatrix} \tilde{p}^{ext}, \quad (22)$$

however, this inversion is only possible if the matrix $[\mathbf{R}]$ is invertible. We will show later that conditions of invertibility agree with the requirements of the stability of the pre-stressed reference state. This is the problem we shall address now.

## 4 Linear stability analysis of pre-stressed reference states

The scope of this section is not to embed the equilibrium properties of our system into a full non linear dynamic stability analysis, we simply wish to know whether a particular given reference state is stable with respect to small perturbations. In this spirit, the reference states described by the density field $\rho_s^0, \rho_f^0$ and the pressure $p_0^{ext}$ will now be assumed to be *spatially uniform*. Our interest is in their stability against linear perturbations of the densities and the pressures. It is expected that such stability properties depend on the functional form of the total energy

$$\psi^{tot} = \psi^{int} - \psi^{ext} \quad (23)$$

and its properties around an equilibrium state. In a static situation stability of the equilibrium state requires the function $\psi^{tot}$ to assume its minimum in equilibrium (Dirichlet criterion), so that the Hessian matrix $\left[\mathfrak{H}\left(\psi^{tot}\right)\right]$ is positive definit in a neighbourhood of the equilibrium,

$$\mathfrak{H} = \begin{bmatrix} 2\gamma_s + \rho_0\gamma_{ss} & \gamma_s + \gamma_f + \rho_0\gamma_{sf} \\ \gamma_s + \gamma_f + \rho_0\gamma_{sf} & 2\gamma_f + \rho_0\gamma_{ff} \end{bmatrix}, \quad (24)$$

implying the Rouse-Hurwitz criteria

$$2\gamma_s + \rho_0\gamma_{ss} > 0, \qquad \det \mathfrak{H} > 0. \quad (25)$$

With the help of $(19)_1$ the firs inequality implies

$$2\left(\frac{p_0^{ext}}{\rho_0^2} + \underbrace{\frac{\rho_f^0}{\rho_0^2}(c_s - c_f)}_{>-1/\rho_f^0} p_0^{ext}\right) + \rho_0\gamma_{ss} > 0,$$

requiring at worst that $\gamma_{ss} > 0$. We will therefore suppose that $\gamma_{ss} > 0$ for all cases. The second inequality for stability can be written as

$$-\rho_0^2 (c_s - c_f)^2 \left(p_0^{ext}\right)^2 + 2\rho_0^3 \beta\, p_0^{ext} + \rho_0^6 \left(\gamma_{ss}\gamma_{ff} - \gamma_{sf}^2\right) > 0, \quad (26)$$

in which

$$\beta := \beta_0 + (c_s - c_f)\beta_1,$$
$$\beta_0 := (\gamma_{ss} + \gamma_{ff} - 2\gamma_{sf}),$$
$$\beta_1 := \left[\rho_s^0(\gamma_{sf} - \gamma_{ss}) + \rho_f^0(\gamma_{ff} - \gamma_{sf})\right]. \quad (27)$$

The LHS of (26) is a quadratic form $P_2(p_0^{ext})$. It represents a set of parabolas (see Fig. 1)[2] with a positive value at the vertex and which are open in the downward direction; the two solutions of the equation $P_2(p_{0*}^{ext}) = 0$,

$$p_{0*}^{ext} = -\frac{1}{(c_s - c_f)^2}\left[-\rho_0\beta \pm \sqrt{\rho_0^2\beta^2 + \rho_0^4(c_s - c_f)^2(\gamma_{ss}\gamma_{ff} - \gamma_{sf}^2)}\right], \quad (28)$$

are positive and negative irrespective of whether $\beta > 0$ or $\beta < 0$. The stability region is $p_0^{ext} < (p_0^{ext})_C$, where $(p_0^{ext})_C$ is the root of (28) with the negative square root sign (as the other root is negative for all admissible choices of the involved parameters).

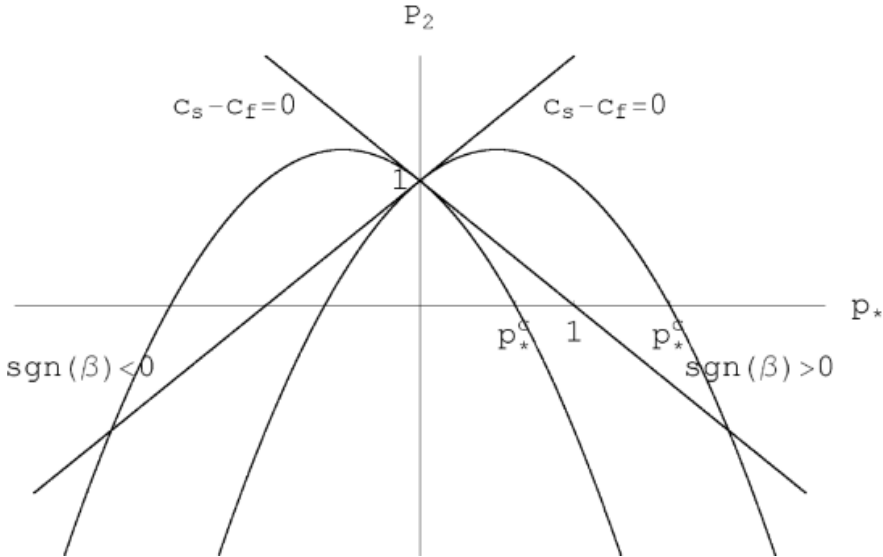

**Fig. 1.** Plot of $P_2(p_*) = 0$ for a value of $\alpha = \pm 1$. Two parabolas are obtained if $c_s - c_f \neq 0$. The parabolas have their vertices at $p_* = 1/2\alpha$ and are open for negative values of $p_*$. If $p_* > p_{*c}$ then the static equilibrium of the basic state is unstable. For $c_s - c_f = 0$ the two parabolas reduce to the same pair of straight lines. The one with positive slope has an unbounded stability limit, while that with negative slope has a finit stability limit. These limits are denoted by $p_{*c}$

When $(c_s - c_f)$ is zero or $[\rho] \to \infty$ then inequality (26) reduces to a linear statement in $p_0^{ext}$ and the parabolas become straight lines. For $\beta_0 > 0$ the stability limit for $p_0^{ext}$ is unbounded, while for $\beta_0 < 0$ it is bounded and is given by

$$(p_0^{ext})_C = \rho_0^3 \frac{\gamma_{ss}\gamma_{ff} - \gamma_{sf}^2}{2(2\gamma_{sf} - \gamma_{ss} - \gamma_{ff})}. \quad (29)$$

---

[2] Figure 1 is a condensed graphical representation in which the transformation

$$p_* = \frac{p_0^{ext}}{\bar{p}}, \quad \bar{p} := \frac{\rho_0^6(\gamma_{ss}\gamma_{ff} - \gamma_{sf}^2)}{2\rho_0^3|\beta|}$$

reduces (26) to

$$P_2(p_*) := -\alpha p_*^2 \pm p_* + 1 > 0, \quad \alpha := \rho_0^2(c_s - c_f)^2 \frac{\rho_0^6(\gamma_{ss}\gamma_{ff} - \gamma_{sf}^2)}{(2\rho_0^3|\beta|)^2},$$

thus collapsing the family of parabolas to two single graphs.

Noticing that

$$\beta_0 > 0 \Rightarrow \frac{\gamma_{ss} + \gamma_{ff}}{2} > \gamma_{sf}, \qquad \beta_0 < 0 \Rightarrow \frac{\gamma_{ss} + \gamma_{ff}}{2} < \gamma_{sf} \qquad (30)$$

we deduce that the firs case corresponds to a weak coupling of the solid and the flui phases via $\gamma_{sf}$, whilst the second one is related to a strong coupling. So stability exists for all pressures in the presence of weak coupling, whilst for strong coupling stability is restricted to small pre-stresses.

These conditions change qualitatively, when $(c_s - c_f)$ is not equal to zero. The stability limit is now bounded in both cases, $\beta > 0$ and $\beta < 0$. Under these circumstances instability is always reached if only $p_0^{ext}$ is sufficientl large. To be more exact equation (28) implies that[3]

$$\left(p_0^{ext}\right)_C = \frac{1}{(c_s - c_f)^2} \left[\rho_0 \beta + \sqrt{\rho_0^2 \beta^2 + \rho_0^4 (c_s - c_f)^2 \left(\gamma_{ss}\gamma_{ff} - \gamma_{sf}^2\right)}\right] \qquad (31)$$

is a positive function of $(c_s - c_f)$, irrespective of whether $\beta > 0$ or $\beta < 0$; it is increasing (decreasing) for negative (positive) values of $(c_s - c_f)$, assumes its maximum at $(c_s - c_f) = 0$ and local minima at the boundaries $|c_s - c_f| = \min(1/\rho_s^0, 1/\rho_f^0)$. Denoting by $(c_s - c_f)_B$ the point at which $\left(p_0^{ext}\right)_C$ attains its minimum, it follows from (31) that the minimum critical pressure $\left(p_0^{ext}\right)_C\big|_{(c_s - c_f)_B}$ depends on, see (27),

$$\beta|_B := \underbrace{\left(1 - \rho_s^0 (c_s - c_f)_B\right)}_{\geq 0} (\gamma_{ss} - \gamma_{sf}) + \underbrace{\left(1 + \rho_f^0 (c_s - c_f)_B\right)}_{\geq 0} (\gamma_{ff} - \gamma_{sf}). \qquad (32)$$

Thus, $\beta|_B$ is a weighted average of $(\gamma_{ss} - \gamma_{sf})$ and $(\gamma_{ff} - \gamma_{sf})$ with weights which depend on the division of the external traction onto the solid and the flui phases. Moreover, $\beta|_B$ is positive (negative) according to whether $(\gamma_{ss} - \gamma_{sf})$ and $(\gamma_{ff} - \gamma_{sf})$ are positive (negative). Since strong coupling corresponds to large $|\gamma_{sf}|$ values, it is evident that it enhances the potential of instability.

Definin

$$I = \frac{\beta|_B}{\rho_0 |(c_s - c_f)_B| \sqrt{\left(\gamma_{ss}\gamma_{ff} - \gamma_{sf}^2\right)}} \qquad (33)$$

for negative $\beta|_B$, (31) takes the form

$$\left(p_0^{ext}\right)_C\big|_{(c_s - c_f)_B} = \frac{\rho_0 \, \beta|_B}{(c_s - c_f)^2} \left[1 + \sqrt{1 + 1/I^2}\right].$$

Therefore, the larger $I^2$, or the more negative $I$ is, the closer to zero will be $\left(p_0^{ext}\right)_C\big|_{(c_s - c_f)_B}$. Consequently $I$ can be taken as a measure of instability.

## 5 Conditions for pressure induced dilatancy of the solid matrix

Notice that our intention is to fin pressurized conditions that yield dilatancy. It is plain, that such states make only sense if they are stable. This was the reason why the stability analysis was presented in the firs place. In this spirit, let us return to the system (20) determining the perturbations $\tilde{\rho}_s$ and $\tilde{\rho}_f$, if $\tilde{p}^{ext}$ is given. A formal inversion of this system of equations yields

$$\tilde{\rho}_s = \frac{\det\left[R_{\rho_s}\right]}{\det[R]} = \frac{\det\left[R_{\rho_s}\right]}{\rho_s^0 \rho_f^0 \det[\mathfrak{H}]}, \qquad (34)$$

$$\tilde{\rho}_f = \frac{\det\left[R_{\rho_f}\right]}{\det[R]} = \frac{\det\left[R_{\rho_f}\right]}{\rho_s^0 \rho_f^0 \det[\mathfrak{H}]},$$

---
[3] We explicitly remark that when $c_s - c_f$ tends to zero expression (31) tends to infinit if $\beta > 0$ but to (29) if $\beta < 0$.

where

$$\det[R_{\rho_s}] = \tilde{p}^{ext} \frac{\rho_s^0 \rho_f^0}{\rho_0^3} \left[ p_0^{ext} \rho_0 (c_s - c_f) \left(1 - (c_s - c_f) \rho_s^0\right) \right. \quad (35)$$
$$\left. + \rho_0^3 \left(1 + (c_s - c_f) \rho_f^0\right) \gamma_{ff} - \rho_0^3 \left(1 - (c_s - c_f) \rho_s^0\right) \gamma_{sf} \right],$$

$$\det[R_{\rho_f}] = \tilde{p}^{ext} \frac{\rho_s^0 \rho_f^0}{\rho_0^3} \left[ -p_0^{ext} \rho_0 (c_s - c_f) \left(1 + (c_s - c_f) \rho_f^0\right) \right.$$
$$\left. + \rho_0^3 \left(1 - (c_s - c_f) \rho_s^0\right) \gamma_{ss} - \rho_0^3 \left(1 + (c_s - c_f) \rho_f^0\right) \gamma_{sf} \right].$$

These can be computed by using Cramer's rule implemented in MAPLE. For stability $\det[\mathfrak{H}] > 0$, so the signs of $\tilde{\rho}_s$, $\tilde{\rho}_f$ are dictated by $\det[R_{\rho_s}]$ and $\det[R_{\rho_f}]$, respectively.

To obtain conditions of dilatancy of the solid matrix, induced by an increase of pressure applied on its boundary, one can remark that $\tilde{\rho}_s$ is negative if and only if $\det[R_{\rho_s}] < 0$, and this requires that

1. if $(c_s - c_f) > 0$,

$$p_0^{ext} < \frac{\rho_0^2 \left[ \left(1 - (c_s - c_f) \rho_s^0\right) \gamma_{sf} - \left(1 + (c_s - c_f) \rho_f^0\right) \gamma_{ff} \right]}{(c_s - c_f)\left(1 - (c_s - c_f) \rho_s^0\right)}, \quad (36)$$

2. if $(c_s - c_f) < 0$,

$$p_0^{ext} > \frac{\rho_0^2 \left[ \left(1 + (c_s - c_f) \rho_f^0\right) \gamma_{ff} - \left(1 - (c_s - c_f) \rho_s^0\right) \gamma_{sf} \right]}{|c_s - c_f|\left(1 - (c_s - c_f) \rho_s^0\right)}, \quad (37)$$

3. if $(c_s - c_f) = 0$, (35)$_1$ implies that $\det[R_{\rho_s}]$ is independent of the pressure $p_0^{ext}$. Then

$$\det[R_{\rho_s}]\Big|_{(c_s - c_f)=0} = \tilde{p}^{ext} \rho_s^0 \rho_f^0 (\gamma_{ff} - \gamma_{sf}), \quad (38)$$

and, therefore, $\tilde{\rho}_s$ is negative provided that

$$\gamma_{ff} < \gamma_{sf}. \quad (39)$$

If the RHS of (36) is negative then a solution with $\tilde{\rho}_s < 0$ does not exist for $p_0^{ext} > 0$. Should the RHS of (37) be negative, then dilatancy occurs for all $p_0^{ext} > 0$.

The foregoing analysis shows that dilatancy under external pressure is possible in a material if only the coupling coefficient $\gamma_{sf}$ is sufficiently large. This, however, does not yet demonstrate that a real material exists such that dilatancy is indeed established.

## 6 Second gradient energy describing pore micro-deformations

The model equations derived so far enjoy the following properties: when a uniform external pressure is applied to the mixture body, exhibiting purely spherical stress states, the constituent densities are equally uniform. This is so because the model does not account for the possible formation of a boundary layer (in which apparent densities are spatially variable) at the mixture external interface. From a physical point of view one could state that this absence is a consequence of a lacking description of the microscopic pore deformation. This is a singular behaviour of the first gradient theory.

To cure the first gradient model from such singular features we now develop a second gradient mixture model. Such theories were preveously developed as non-simple mixture models (Müller [16], Rajagopal & Tao [17]). We are not aware of any second gradient PVP-approach for mixtures in the spirit pursued here, but our approach essentially follows Gouin [13], Casal [3] and Seppecher [20] (for simple mixtures) and Germain [12] who use the PVP for single constituent bodies.

*6.1 General balance equations*

Let us now consider the PVP for the general situation in which (i) body forces $\mathbf{b}_a$ are present, (ii) the internal energy $\psi^{int}$ may also depend on the second deformation gradients and (iii) the action of the exterior to the body is given by tractions $\mathbf{t}_a$ and double forces $\tau_a$ (see for more details [12] and [9]). For such a case equation (5) is generalized in the form

$$\frac{d}{dt}\int_{\mathcal{B}} \psi^{int} dV = \int_{\mathcal{B}} \mathbf{b}_a \cdot \mathbf{v}_a \, dV + \int_{\partial\mathcal{B}} \left(\mathbf{t}_a \cdot \mathbf{v}_a + \tau_a \cdot \frac{\partial \mathbf{v}_a}{\partial \mathbf{n}}\right) dA; \tag{40}$$

$\tau_s$ and $\tau_f$ are the double forces acting on the solid and on the fluid respectively. We select the simplest gradient dependence of $\psi^{int}$,

$$\psi^{int} = \varepsilon(\rho_a) + \frac{\lambda_s}{2} f_{ss}, \qquad f_{ss} := \nabla \rho_s \cdot \nabla \rho_s, \tag{41}$$

where $\lambda_s$ is a constant. This corresponds to a gradient dependence for the solid but not for the fluid which is special. The localization of (40) with $\lambda_s = 0$ is easily shown to be

$$\nabla p_a - \mathbf{m}_a = \mathbf{b}_a, \quad \text{in } \mathcal{B}, \qquad p_a = d_a p^{ext}, \quad \text{on } \partial\mathcal{B}, \tag{42}$$

where

$$\begin{aligned} p_a &:= \rho_a \frac{\partial \varepsilon}{\partial \rho_a} - \xi_a \varepsilon, \\ \mathbf{m}_a &:= \frac{\partial \varepsilon}{\partial \rho_a} \nabla \rho_a - \nabla(\xi_a \varepsilon), \qquad \text{(no sum. over } a) \end{aligned} \tag{43}$$

The details of the derivation follow eqs.(7) (8) (9) and are e.g. also given in [9]. It therefore suffice do deal with the additional gradient dependent term in (40)

$$\begin{aligned} I_{add} &:= \left(\frac{d}{dt}\int_{\mathcal{B}} \psi^{int} dV\right)_{add} \\ &= \frac{\lambda_s}{2}\frac{d}{dt}\int_{\mathcal{B}} \nabla\rho_s \cdot \nabla\rho_s \, dV = \int_{\partial\mathcal{B}} \left(\mathbf{t}_{a_{add}} \cdot \mathbf{v}_a + \tau_a \cdot \frac{\partial \mathbf{v}_a}{\partial \mathbf{n}}\right). \end{aligned} \tag{44}$$

Applying the Reynolds Transport Theorem yields (see Appendix A)

$$\begin{aligned} I_{add} &= \lambda_s \int_{\mathcal{B}} \left[\frac{1}{2}(f_{ss})\mathbf{I} \cdot \nabla(\xi_a \mathbf{v}_a) + \nabla\rho_s \cdot \left(\frac{\partial}{\partial t}\nabla\rho_s + \nabla \otimes \nabla\rho_s(\xi_a \mathbf{v}_a)\right)\right] dV \\ &= -\lambda_s \int_{\mathcal{B}} \left[div\left(\frac{1}{2}f_{ss}\xi_s\mathbf{I} - f_{ss}\mathbf{I} - \nabla\rho_s \otimes \nabla\rho_s + div(\rho_s \nabla\rho_s \otimes \mathbf{I})\right)\right.\\ &\qquad\left. + \frac{1}{2}\nabla f_{ss} - \frac{1}{2}\nabla(\xi_s f_{ss})\right] \cdot \mathbf{v}_s \, dV \\ &\quad + \lambda_s \int_{\partial\mathcal{B}} \left\{\left[\frac{1}{2}f_{ss}\xi_s\mathbf{I} - f_{ss}\mathbf{I} - \nabla\rho_s \otimes \nabla\rho_s + div(\rho_s \nabla\rho_s \otimes \mathbf{I})\right]\mathbf{n} \cdot \mathbf{v}_s \right.\\ &\qquad \left. + \frac{1}{2}f_{ss}\xi_f \mathbf{n} \cdot \mathbf{v}_f - (\rho_s \nabla\rho_s \otimes \mathbf{I})\mathbf{n} \cdot \nabla\mathbf{v}_s \right\} dA. \end{aligned} \tag{45}$$

Consequently, localization of the complete equation (40) leads to the following boundary value problem:

$$\begin{aligned} \nabla p_s - \mathbf{m}_s - \lambda_s \, div\, \left(\rho_s \triangle \rho_s \mathbf{I} + \tfrac{1}{2}f_{ss}\mathbf{I} - \nabla\rho_s \otimes \nabla\rho_s\right) &= \mathbf{b}_s, \\ &\qquad \text{in } \mathcal{B} \\ \nabla p_f - \mathbf{m}_f &= \mathbf{b}_f, \end{aligned} \tag{46}$$

$$-p_s \mathbf{n} + \lambda_s \left(\rho_s \triangle \rho_s \mathbf{I} + \tfrac{1}{2} f_{ss} \xi_s \mathbf{I} - \nabla \rho_s \otimes \nabla \rho_s\right) \mathbf{n} = \mathbf{t}_s,$$

$$-p_f \mathbf{n} + \frac{\lambda_s}{2} f_{ss} \xi_f \mathbf{n} = \mathbf{t}_f,$$

$$-\lambda_s \rho_s \left(\nabla \rho_s \cdot \mathbf{n}\right) \mathbf{n} = \tau_s,$$

$$\mathbf{0} = \tau_f,$$

on $\partial \mathcal{B}$. (47)

Notice that the gradient effects only enter the field equations of the solid, this obviously because of our restrictive assumption (41). The boundary conditions of the fluid are, however, affected by the gradient terms; they generate an additional pressure. The surface double forces only enter the boundary conditions of the solid constituent, because the free energy does not depend on $\nabla \rho_f$.

*6.2 External action potential*

We now address the problem of finding a potential $\psi^{ext}$ for the external action specified on RHS of (40). We are looking for a potential $\psi^{ext}$ that depends on the state parameters $\rho_s$, $\rho_f$, and $\nabla \rho_s$, such that

$$\frac{d}{dt} \int_{\mathcal{B}} \psi^{ext} dV = \int_{\mathcal{B}} \mathbf{b}_a \cdot \mathbf{v}_a \, dV + \int_{\partial \mathcal{B}} \left(\mathbf{t}_a \cdot \mathbf{v}_a + \tau_a \cdot \frac{\partial \mathbf{v}_a}{\partial \mathbf{n}}\right) dA. \tag{48}$$

Its existence assures that a cyclic quasi-static variation of these parameters is path-independent. Using the Reynolds Transport Theorem the LHS of (48) becomes

$$\frac{d}{dt} \int_{\mathcal{B}} \psi^{ext} dV = \int_{\mathcal{B}} \left[\frac{\partial}{\partial t} \psi^{ext} + div\left(\psi^{ext} \mathbf{v}\right)\right] dV. \tag{49}$$

Performing the differentiations term by term and using the constituent balances of mass yields

$$\frac{\partial}{\partial t} \psi^{ext} = -\frac{\partial \psi^{ext}}{\partial \rho_a} \rho_a \mathbf{I} \cdot \nabla \mathbf{v}_a - \frac{\partial \psi^{ext}}{\partial \rho_a} \nabla \rho_a \cdot \mathbf{v}_a - \left(\frac{\partial \psi^{ext}}{\partial \nabla \rho_s} \cdot \nabla \rho_s\right) \mathbf{I} \cdot \nabla \mathbf{v}_s$$

$$-\nabla \otimes \nabla \rho_s \frac{\partial \psi^{ext}}{\partial \nabla \rho_s} \cdot \mathbf{v}_s + \left(\frac{\partial \psi^{ext}}{\partial \nabla \rho_s} \otimes \nabla \rho_s\right) \cdot \nabla \mathbf{v}_s$$

$$-\left(\rho_s \frac{\partial \psi^{ext}}{\partial \nabla \rho_s} \otimes \mathbf{I}\right) \cdot \nabla \otimes \nabla \mathbf{v}_s \tag{50}$$

$$div\left(\psi^{ext} \mathbf{v}\right) = div\left(\psi^{ext} \xi_a \mathbf{v}_a\right) = \xi_a \psi^{ext} div \, \mathbf{v}_a + \mathbf{v}_a \cdot \nabla \left(\xi_a \psi^{ext}\right).$$

Substituting these above allows to write (49) as

$$\frac{d}{dt} \int_{\mathcal{B}} \psi^{ext} dV = \int_{\mathcal{B}} \left(\beta_a \cdot \mathbf{v}_a + \mathbf{B}_a \cdot \nabla \mathbf{v}_a + \mathfrak{B}_s \cdot \nabla \otimes \nabla \mathbf{v}_s\right) dV, \tag{51}$$

where the quantities $\beta_a$, $\mathbf{B}_a$ ($a = s, f$) and $\mathfrak{B}_s$ are defined by the following expressions:

$$\beta_s := -\frac{\partial \psi^{ext}}{\partial \rho_s} \nabla \rho_s - (\nabla \otimes \nabla \rho_s) \frac{\partial \psi^{ext}}{\partial (\nabla \rho_s)} + \nabla(\xi_s \psi^{ext}),$$

$$\beta_f := -\frac{\partial \psi^{ext}}{\partial \rho_f} \nabla \rho_f + \nabla(\xi_f \psi^{ext}),$$

$$\mathbf{B}_s := \left[-\frac{\partial \psi^{ext}}{\partial \rho_s} \rho_s - \left(\frac{\partial \psi^{ext}}{\partial (\nabla \rho_s)} \cdot \nabla \rho_s\right) + \xi_s \psi^{ext}\right] \mathbf{I} - \frac{\partial \psi^{ext}}{\partial (\nabla \rho_s)} \otimes \nabla \rho_s, \tag{52}$$

$$\mathbf{B}_f := \left(-\frac{\partial \psi^{ext}}{\partial \rho_f} \rho_f + \xi_f \psi^{ext}\right) \mathbf{I},$$

$$\mathfrak{B}_s := -\rho_s \frac{\partial \psi^{ext}}{\partial (\nabla \rho_s)} \otimes \mathbf{I}.$$

The RHS of (48) and (51) agree with one another if $\mathbf{b}_a$, $\mathbf{t}_a$, and $\boldsymbol{\tau}_a$ are given by

$$\begin{aligned}
\mathbf{b}_s &= \boldsymbol{\beta}_s - div\,(\mathbf{B}_s - div\,\mathfrak{B}_s), & \mathbf{b}_f &= \boldsymbol{\beta}_f - div\,\mathbf{B}_f, \\
\mathbf{t}_s &= (\mathbf{B}_s - div\,\mathfrak{B}_s)\mathbf{n} - div_s(\mathfrak{B}_s\mathbf{n}), & \mathbf{t}_f &= \mathbf{B}_f\mathbf{n}, \\
\boldsymbol{\tau}_s &= (\mathfrak{B}_s\mathbf{n})\mathbf{n}, & \boldsymbol{\tau}_f &= \mathbf{0};
\end{aligned} \tag{53}$$

in these formulas we consider the following decomposition for the gradient of an $n$-th order tensor field $\Omega$:

$$\nabla\Omega = \nabla^s\Omega + \mathbf{n}\otimes\frac{\partial\Omega}{\partial\mathbf{n}}, \tag{54}$$

$\nabla^s\Omega$ being the restriction of $\nabla\Omega$ on $\partial\mathcal{B}$, and we define the *surface divergence* of an $n$-th order tensor field $\Omega$, the differential operator $div_s$ such that

$$div_s\left(\Omega^T u\right) = (div_s\,\Omega)\cdot u + \Omega\cdot\nabla^s u, \quad \forall (n-1)\text{-th order tensor field } u \tag{55}$$

and

$$\int_{\partial\mathcal{B}} div_s\,\Omega\,dA = \int_{\partial\partial\mathcal{B}} \Omega\nu\,dS, \tag{56}$$

where $\nu$ is the outward normal to $\partial\partial\mathcal{B}$ the line boundary of $\partial\mathcal{B}$. For smooth surfaces $\partial\mathcal{B}$ the integral on the RHS of (56) vanishes. In this case the contact action on the solid and the fluid the double forces on the solid and the fluid are given by (see Appendix B)

$$\begin{aligned}
\mathbf{t}_s &= \left[-\frac{\partial\psi^{ext}}{\partial\rho_s}\rho_s + \xi_s\psi^{ext} - \frac{\partial\psi^{ext}}{\partial(\nabla\rho_s)}\cdot\nabla\rho_s + div\left(\rho_s\frac{\partial\psi^{ext}}{\partial(\nabla\rho_s)}\right)\right.\\
&\quad \left.+\rho_s\frac{\partial\psi^{ext}}{\partial(\nabla\rho_s)}\cdot\mathbf{n}\,(tr\,\nabla^s\mathbf{n}) - \left(\frac{\partial\psi^{ext}}{\partial(\nabla\rho_s)}\cdot\mathbf{n}\right)\frac{\partial\rho_s}{\partial\mathbf{n}}\right]\mathbf{n}+\rho_s\nabla^s\left(\frac{\partial\psi^{ext}}{\partial(\nabla\rho_s)}\cdot\mathbf{n}\right), \\
\mathbf{t}_f &= \left(-\frac{\partial\psi^{ext}}{\partial\rho_f}\rho_f + \xi_f\psi^{ext}\right)\mathbf{n}, \\
\boldsymbol{\tau}_s &= -\left(\rho_s\frac{\partial\psi^{ext}}{\partial(\nabla\rho_s)}\cdot\mathbf{n}\right)\mathbf{n}, \\
\boldsymbol{\tau}_f &= \mathbf{0}.
\end{aligned} \tag{57}$$

We distinguish the normal and the shear parts of $\mathbf{t}_s$ and assume that the double force acting on the solid depends linearly on the external pressure: $\boldsymbol{\tau}_s = d_D\,p^{ext}\mathbf{n}$; this is reasonable, since increasing the pressure increases the pore space and the latter is opened by the action of the double force. In so doing we obtain the following forms of the constitutive relations for the coefficient $d_D$ and for the coefficient $d_a$ ($a=s,f$), appearing in (4) valid in the case of second gradient solid matrices:

$$\begin{aligned}
d_D &= -\frac{1}{p^{ext}}\left(\rho_s\frac{\partial\psi^{ext}}{\partial(\nabla\rho_s)}\cdot\mathbf{n}\right), \\
d_s &= \frac{1}{p^{ext}}\left[\frac{\partial\psi^{ext}}{\partial\rho_s}\rho_s - \xi_s\psi^{ext} - \rho_s\,div\left(\frac{\partial\psi^{ext}}{\partial\nabla\rho_s}\right)\right.\\
&\quad \left.+\left(\rho_s\frac{\partial\psi^{ext}}{\partial\nabla\rho_s}\cdot\mathbf{n}\right)tr\,\nabla^s\mathbf{n} + \left(\frac{\partial\psi^{ext}}{\partial\nabla\rho_s}\cdot\mathbf{n}\right)\nabla\rho_s\cdot\mathbf{n}\right], \\
d_f &= \frac{1}{p^{ext}}\left[\frac{\partial\psi^{ext}}{\partial\rho_f}\rho_f - \xi_f\psi^{ext}\right].
\end{aligned} \tag{58}$$

These formulas simply emerge if one divides $\boldsymbol{\tau}_s$, and the components of $\mathbf{t}_a$ normal to the surface by $p^{ext}$.

Assume that the body forces $\mathbf{b}_s$ and $\mathbf{b}_f$ vanish; then equations (53)$_{1,2}$ yield the conditions

$$\rho_s\nabla\left(\frac{\partial\psi^{ext}}{\partial\rho_s} - div\left(\frac{\partial\psi^{ext}}{\partial\nabla\rho_s}\right)\right) = 0, \quad \rho_f\nabla\left(\frac{\partial\psi^{ext}}{\partial\rho_f}\right) = 0. \tag{59}$$

Further investigations will be necessary to generalize the results, found in the §2, about $\psi^{ext}$ implied by the condition $d_f + d_s = 1$. We simply remark here that one can find in the subsequent one-dimensional problem, developed as an application of the introduced new model, a form for $\psi^{ext}$ verifying the above constraints.

## 7 A one dimensional application

Consider a one-dimensional problem in which the body forces on the solid and on the fluid vanish. Assume constant external pressure and suppose that the derivative of $\psi^{ext}$ with respect to $\nabla \rho_s$ is constant in $\mathscr{B}$ i.e.

$$\frac{d}{dx}\left(\frac{\partial \psi^{ext}}{\partial (\rho_{s,x})}\right) = 0, \tag{60}$$

where $\partial \psi^{ext}/\partial (\rho_{s,x}) := \partial \psi^{ext}/\partial \nabla \rho_s \cdot \mathbf{e}$, where $\mathbf{e}$ is the unit vector defining the $x$ direction. With this, eqs. (57) reduce to

$$d_s p^{ext} = \frac{\partial \psi^{ext}}{\partial \rho_s}\rho_s - \xi_s \psi^{ext} + \left(\frac{\partial \psi^{ext}}{\partial (\rho_{s,x})}\right)\frac{d\rho_s}{dx}, \quad d_f p^{ext} = \frac{\partial \psi^{ext}}{\partial \rho_f}\rho_f - \xi_f \psi^{ext},$$

$$\mathbf{t}_s^{shear} = \mathbf{0}, \qquad\qquad\qquad\qquad\qquad \mathbf{t}_f^{shear} = \mathbf{0}, \tag{61}$$

$$\tau_s = -\rho_s \left(\frac{\partial \psi^{ext}}{\partial (\rho_{s,x})}\right)\mathbf{e}, \qquad\qquad \tau_f = \mathbf{0}$$

and equations (59) become

$$\frac{d}{dx}\left(\frac{\partial \psi^{ext}}{\partial \rho_a}\right) = 0, \qquad a = s, f. \tag{62}$$

Eqs. (60) and (62) imply that $\psi^{ext}$ is a linear function of its arguments $\rho_a$ and $d\rho_s/dx$. Thus, with an appropriate normalization, one has

$$\psi^{ext} = p^{ext}\left(c_s \rho_s + c_f \rho_f + k_s \frac{d\rho_s}{dx} - 1\right), \tag{63}$$

whilst the constitutive relations defining the coefficient $d_a$ ($a = s, f$) and $d_D$ are

$$\begin{aligned}
d_s &= \xi_s \left(1 + (c_s - c_f)\rho_f\right) + \xi_f k_s \frac{d\rho_s}{dx}, \\
d_f &= \xi_f \left(1 - (c_s - c_f)\rho_s\right) - \xi_f k_s \frac{d\rho_s}{dx}, \\
d_D &= -k_s \rho_s (\mathbf{n} \cdot \mathbf{e}),
\end{aligned} \tag{64}$$

as easily deducible from (63) and (58) or from (61). Note that constitutive relations for $d_s$ and $d_f$ differ from relations (11), obtained by a first gradient mixture model, by an additive quantity; this is due to the assumption on the derivative of $\psi^{ext}$ with respect to $\nabla \rho_s$: $\psi^{ext}$ simply depends linearly on $\rho_s$ and $d\rho_s/dx$.

Consider a linearized theory and assume that in the reference configuration the constituent density of the solid (and of the fluid) is not uniform. This hypothesis is necessary to appreciate second gradient effects: if the densities were uniform in the reference configuration then $(47)_3$ would imply $\tau_s = 0$, i.e., $k_s = 0$, so that there would be no possibility to have a non-vanishing double force on $\partial \mathscr{B}$, acting on the solid skeleton.

We also assume that the coefficient $\gamma_s$ of the linear term of the internal potential energy $\varepsilon$ is not constant but a linear function of $\rho_s^0$,

$$\gamma_s = \alpha_s \rho_s^0, \tag{65}$$

where $\alpha_s$ is assumed uniform in $\mathscr{B}$; a justification for this will be given shortly. With these prerequisites we may now perform a perturbation analysis in the vicinity of a pre-stressed reference state with $\psi^{ext}$ given by (63) and $\psi^{int}$ by (41). With an approach entirely analogous to that of §3 we then deduce from the balance laws the following zeroth and first order equations:

*zeroth order problem*

$$\frac{d\rho^0}{dx}\alpha_s \rho_s^0 + \rho^0 \alpha_s \frac{d\rho_s^0}{dx} - \lambda_s \frac{d^3 \rho_s^0}{dx^3} = 0, \quad \rho_f^0 \frac{d\rho^0}{dx}\gamma_f = 0, \tag{66}$$

*first order problem*

$$-\lambda_s \frac{d^3 \tilde{\rho}_s}{dx^3} + \left(2\alpha_s \rho_s^0 + \rho^0 \gamma_{ss}\right) \frac{d\tilde{\rho}_s}{dx} + \left(\alpha_s \rho_s^0 + \gamma_f + \rho^0 \gamma_{sf}\right) \frac{d\tilde{\rho}_f}{dx} +$$
$$+ 2\alpha_s \frac{d\rho_s^0}{dx} \tilde{\rho}_s + \alpha_s \frac{d\rho_s^0}{dx} \tilde{\rho}_f = 0, \qquad (67)$$
$$\left(2\gamma_f + \rho^0 \gamma_{ff}\right) \frac{d\tilde{\rho}_f}{dx} + \left(\alpha_s \rho_s^0 + \gamma_f + \rho^0 \gamma_{sf}\right) \frac{d\tilde{\rho}_s}{dx} + \alpha_s \frac{d\rho_s^0}{dx} \tilde{\rho}_s = 0.$$

To these ODEs at each perturbation order four boundary conditions must be added: we suppose that the tractions on the solid and on the fluid and the double force are known at $x = 0$, and we assume that the double force (and therefore $d\rho_s^0/dx$) vanishes as $x \to \infty$. The condition (65) corresponds to the idea that the apparent density of the solid (and of the fluid in the reference state is given by the sum of a constant and an exponentially decreasing term, smaller than zero; if $\gamma_s$ were constant in $\mathscr{B}$ then equation (66)$_1$ would imply that $\rho_s^0 = const$ because of the condition as $x \to \infty$.

The solution of the zeroth order problem is

$$\rho_s^0(x) = C_1 + C_2 \exp\left(\frac{x}{x_0}\right) + C_3 \exp\left(-\frac{x}{x_0}\right),$$
$$\rho^0(x) = C_4,$$

where

$$x_0 := \sqrt{\lambda_s / \rho^0 \alpha_s}. \qquad (68)$$

This explicitly demonstrates that the double forces are responsable for the exponential decay of $\rho_s^0(x)$ as one moves away from the mixture surrounding environment, since $\lambda_s \neq 0$. The boundary condition at $x \to \infty$ implies that $C_2 = 0$; the boundary condition on the value of the double force at $x = 0$ implies that $C_3 \leq 0$, so the apparent densities of the two constituents are given by

$$\rho_s^0(x) = C_1 - |C_3| \exp\left(-\frac{x}{x_0}\right), \qquad (69)$$
$$\rho_f^0(x) = (C_4 - C_1) + |C_3| \exp\left(-\frac{x}{x_0}\right);$$

$x_0$ is the characteristic decay length of the zeroth order solution. We do not show the explicit expression for $C_1$, $(C_4 - C_1)$ and $|C_3|$ which are rather cumbersome; we simply recall that they depend on the external actions and on the interface constitutive parameters (e.g. $c_s - c_f$) and can be interpreted respectively as the apparent solid and fluid mass densities far from the mixture-surrounding environment and their maximum variations induced by applied external double forces.

To compute the solution of the first order problem, consider the following non-dimensionalization of the independent and dependent variables:

$$\xi := \frac{x}{x_0}, \quad r_s := \frac{\tilde{\rho}_s}{C_1}, \quad r_f := \frac{\tilde{\rho}_f}{C_4 - C_1}.$$

Equations (67) constitute a fourth order differential problem; so it can be expressed as a system of first order differential equations given in the following form:

$$\frac{d\mathbf{Y}}{d\xi} = (\mathbf{R} + \exp(-\xi)\mathbf{A}_0 + \exp(-2\xi)\mathbf{A}_1)\mathbf{Y}, \qquad (70)$$

where $\mathbf{Y}$, $\mathbf{R}$, $\mathbf{A}_0$, $\mathbf{A}_1$ are defined by

$$\mathbf{Y} := \begin{pmatrix} r_s \\ r_f \\ dr_s/d\xi \\ d^2 r_s/d\xi^2 \end{pmatrix}, \quad \mathbf{R} := \begin{pmatrix} 0 & 0 & 1 & 0 \\ 0 & 0 & -a_4/a_6 & 0 \\ 0 & 0 & 0 & 1 \\ 0 & 0 & a_1 - a_3 a_4 & 0 \end{pmatrix},$$

$$\mathbf{A}_0 := \begin{pmatrix} 0 & 0 & 0 & 0 \\ -a_5/a_6 & 0 & a_5/a_6 & 0 \\ 0 & 0 & 0 & 0 \\ 2a_2 - a_3 a_5 & a_2 a_6 & 2(a_3 a_5 - a_2) & 0 \end{pmatrix}, \tag{71}$$

$$\mathbf{A}_1 := \begin{pmatrix} 0 & 0 & 0 & 0 \\ 0 & 0 & 0 & 0 \\ 0 & 0 & 0 & 0 \\ a_2 a_5 & 0 & -a_2 a_5 & 0 \end{pmatrix},$$

and

$$\begin{aligned}
a_1 &:= (x_0^2/\lambda_s)(2\alpha_s C_1 + C_4 \gamma_{ss}), \\
a_2 &:= \alpha_s |C_3| x_0^2/\lambda_s, \\
a_3 &:= (x_0^2/\lambda_s)(\alpha_s C_1 + \gamma_f + C_4 \gamma_{sf}), \\
a_4 &:= (\alpha_s C_1 + \gamma_f + C_4 \gamma_{sf})/(2\gamma_f + C_4 \gamma_{ff}), \\
a_5 &:= \alpha_s |C_3|/(2\gamma_f + C_4 \gamma_{ff}), \\
a_6 &:= (C_4 - C_1)/C_1.
\end{aligned} \tag{72}$$

Consider the following change of variable: $z = \exp(-\xi)$; the differential problem (70) becomes

$$\frac{d\mathbf{Y}}{dz} = -\left(\frac{1}{z}\mathbf{R} + \mathbf{A}_0 + z\,\mathbf{A}_1\right)\mathbf{Y}. \tag{73}$$

This change of variable maps the open set $(0, \infty)$ onto the open set $(0, 1)$; boundary conditions in $\xi = 0$ now are given at $z = 1$ and conditions at $\xi \to \infty$ at $z = 0$.

The matrix $\mathbf{A}(z) = -(z^{-1}\mathbf{R} + \mathbf{A}_0 + z\,\mathbf{A}_1)$ has at most a pole at $z = 0$ but it is analytic for $0 < |z| < a$, $a > 0$ and the point $z = 0$ is a *singular point of the first kind* for the system (73) (see [4]), so it fulfill the hypotheses of theorem 3.1, p. 117 and 4.1, p.119 in [4]. Therefore, the fundamental matrix of system (73) is represented in terms of a series, convergent in the set $0 < |z| < a$,

$$\Phi(z) = \left(\sum_{i=0}^{\infty} \mathbf{Q}_i z^i\right) e^{(\ln z)\mathbf{J}} \tag{74}$$

where $\mathbf{J}$ is the canonical form[4] of $\mathbf{R}$, if and only if $\mathbf{R}$ has characteristic roots which do not differ by positive integers, and

$$\begin{aligned}
\mathbf{R}\mathbf{Q}_0 &= \mathbf{Q}_0 \mathbf{J}, \\
\mathbf{Q}_{m+1}[\mathbf{J} + (m+1)\mathbf{I}] &= \mathbf{R}\mathbf{Q}_{m+1} + \sum_{k=0}^{m} \mathbf{A}_k \mathbf{Q}_{m-k},
\end{aligned} \tag{75}$$

where, in this context, $\mathbf{I}$ is the $(4 \times 4)$ unit matrix. So it follows that the fundamental matrix given as a function of $\xi$ is

$$\Phi(\xi) = (\mathbf{Q}_0 + \exp(-\xi)\mathbf{Q}_1 + \exp(-2\xi)\mathbf{Q}_2 + \ldots)\, e^{-\xi \mathbf{J}}. \tag{76}$$

---
[4] The canonical form of a matrix is define as its Jordan form. If the matrix admits linearly independent eigenvectors its canonical form is diagonal.

When truncating this series at the first order term the non-dimensional solid and fluid densities are given by (see Fig. 2)

$$\begin{aligned} r_s &= k_2 + k_1 \left[ \frac{1}{\xi_0} \exp(-\xi) + \left(1 + \frac{1}{\xi_0}\right) \exp\left(-\frac{\xi}{\xi_0}\right) \right], \\ r_f &= k_3 - k_1 \frac{a_4}{a_6} \left[ \frac{1}{\xi_0} \exp(-\xi) + \left(1 + \frac{1}{\xi_0}\right) \exp\left(-\frac{\xi}{\xi_0}\right) \right], \end{aligned} \tag{77}$$

where $\xi_0 := (a_1 - a_3 a_4)^{-1/2}$, $k_i$ are integration constants to be determined by imposing the boundary conditions implied by (61) at the first order; experiments must give information on their values.

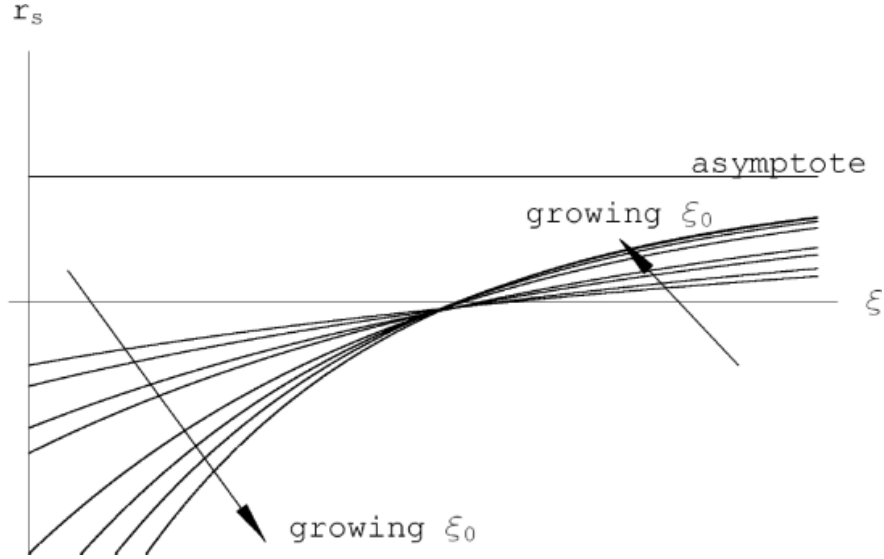

**Fig. 2.** Scaled solid density $r_s$ plotted against dimensionless distance $\xi$ parameterized for various values of the e-folding distance $\xi_0$. All curves approach the asymptote as $\xi \to \infty$

With the above expressions for $a_j$ ($j = 1, 3, 4$), we have

$$\sqrt{a_1 - a_3 a_4} = \frac{x_0}{\sqrt{\lambda_s}} \sqrt{\frac{(2\gamma_f + C_4 \gamma_{ff})(2\alpha_s C_1 + C_4 \gamma_{ss}) - (\alpha_s C_1 + \gamma_f + C_4 \gamma_{sf})^2}{2\gamma_f + C_4 \gamma_{ff}}}. \tag{78}$$

Considering that $C_4 = \rho^0$ and in a first gradient theory $C_1 = \rho_s^0$ (compare with (69)) we may rewrite (78) on using (65) as

$$\sqrt{a_1 - a_3 a_4} = \frac{x_0}{\sqrt{\lambda_s}} \sqrt{\frac{\det \mathfrak{H}}{\mathfrak{H}_{22}}}, \tag{79}$$

where $\mathfrak{H}$ is defined in (24). Thus

$$\xi_0 = \frac{\sqrt{\lambda_s}}{x_0} \sqrt{\frac{\mathfrak{H}_{22}}{\det \mathfrak{H}}}. \tag{80}$$

The eigenvalues of $\mathbf{R}$ are

$$\left\{ \pm \sqrt{a_1 - a_3 a_4} = \pm \xi_0^{-2}, 0, 0 \right\}.$$

In the stability regime one has $\det \mathfrak{H} > 0$, and $\mathfrak{H}_{22} > 0$; therefore, $\xi_0$ is real valued. The fundamental solution of (73) corresponding to the negative eigenvalue of $\mathbf{R}$ grows with increasing $\xi$, so the boundary condition at $\xi \to \infty$ inforcing regularity requires this solution to be absent in (77).

The parameter $\xi_0$ can be greater than unity, if

$$\det \mathfrak{H} = \mathfrak{H}_{11}\mathfrak{H}_{22} - \mathfrak{H}_{12}^2 < \left(\rho^0 \alpha_s\right) \mathfrak{H}_{22}, \tag{81}$$

and it will eventually tend to infinity when $\mathfrak{H}_{12}^2 \to \mathfrak{H}_{11}\mathfrak{H}_{22}$. This last condition occurs if the coupling $\gamma_{sf}$ is sufficiently large: the effect of the second-gradient-depending-deformation energy is an increasing widening of the boundary layer near the external surface of the body when the instability conditions are approached.

## 8 Conclusions

In this paper a binary mixture model was presented which possesses the ingredients that an external pressure may cause a dilatation of the pores, a phenomenon sometimes observed in heterogeneous porous materials. Terzaghi and in particular Fillunger were aware of this phenomenon and knew that their models could not predict this behaviour, and they dismissed the possibility after extensive search for evidence and own experimentation[5]. Perhaps they were too ambitious, for their arguments seem to suggest that they were looking for an explosion of the solid matrix of a pressurized solid-fluid mixture if only the external pressure would be sufficiently large.

The present paper showed within the context of a very simple mixture model – too simple to describe the deformation of the solid accurately, but sufficiently complex to isolate this detail – how Terzaghi's and Fillunger's search could be interpreted. To this end we assumed all constitutive quantities to depend on the apparent densities of the solid and the fluid (and eventually on their gradients) and no more. It turned out that a dependence of the internal free energy on the interaction term $\rho_s \rho_f$, i.e., on the product of the apparent solid and fluid densities, is important. Terzaghi's and Fillunger's *explosion* is interpreted here as the loss of stability of a pre-stressed reference state. The critical external pressure, which causes this reference state to become unstable, is dictated by two physically distinct properties: i) the coefficients of the quadratic terms of the free internal energy and in particular its interaction term and ii) the parameterization how the external pressure is distributed between the solid and the fluid normal boundary tractions. There are parameter sets for which instability never arises and others for which instability can, in principle, always occur, if only the external pressure is sufficiently large (§4).

The second question addressed by Terzaghi and Fillunger is the mentioned dilatancy phenomenon viewed possibly by them to be the same as the *explosive* problem of the solid matrix. Our model also gives an answer to this question. If the pressure corresponding to our reference state is increased then the new stable state can exist only under the above mentioned stability conditions. This new state possesses a smaller or larger apparent solid density provided that the interaction term of the free energy obeys certain equalities involving the other coefficients of the free energy and the parametrization of the constituent boundary tractions (§5). We find it most intriguing, that thermodynamics of bulk and boundary quantities provides the answer to this subtle behaviour of the mixture.

While we do now understand how the above mentioned dilatational effect can be predicted by the model equations, it does not have typical boundary layer structure in a first gradient theoretical setting. This boundary enhancement can be achieved by adding a density-gradient dependence of the solid phase to the free energy [9]. This dependence will lead to an enhancement of the pore space close to the interface between the mixture and the exterior world which dies out as one moves away from the interface, (§6).

The one-dimensional problem which we solve in the last section (§7) proves that second gradient regularization is necessary if one wants to describe the *explosion* phenomenon imagined by Fillunger and Terzaghi. Indeed if one interprets it as a loss of stability the transition from stable to unstable states can (in the framework of the present model) be parameterized by the coupling coefficient $\gamma_{sf}$, *ceteris paribus*. When $\gamma_{sf}$ is increased, $\mathfrak{H}_{12}^2$ also increases and $\det \mathfrak{H}$ tends to zero: the boundary layer at the interface between the mixture body and the external world becomes wider and wider and eventually occupies the whole body before the instability conditions arise. As the second gradient boundary layer is characterized by a lower value of the apparent solid mass density one can state that instability is attained by a progressive dilatational process which is induced by the pore fluid pressure and initially arises at the boundary of the mixture body.

---

[5] As beautifully summarized by [7], Fillunger dismissed the fact that the pore pressure would affect the strength of the porous material.

**Appendix A**

In this section we give the details of the calculation which permit us to obtain relation (45). Using the balance of mass for the solid gives

$$\begin{aligned}\frac{1}{\lambda_s}I_{add} &= \int_{\mathcal{B}}\left[\frac{1}{2}f_{ss}\mathbf{I}\cdot\nabla\left(\xi_a\mathbf{v}_a\right)+\nabla\rho_s\cdot\left(\frac{\partial}{\partial t}\nabla\rho_s+\nabla\otimes\nabla\rho_s\left(\xi_a\mathbf{v}_a\right)\right)\right]\,dV \\ &= \int_{\mathcal{B}}\left[\frac{1}{2}f_{ss}\mathbf{I}\cdot\left(\xi_a\nabla\mathbf{v}_a+\nabla\xi_a\otimes\mathbf{v}_a\right)-\nabla\rho_s\cdot\rho_s\left(\nabla\otimes\nabla\mathbf{v}_s\right)^T\mathbf{I}\right. \\ &\quad -f_{ss}\mathbf{I}\cdot\nabla\mathbf{v}_s-\nabla\mathbf{v}_s\cdot\nabla\rho_s\otimes\nabla\rho_s-\nabla\rho_s\cdot\left(\nabla\otimes\nabla\rho_s\right)\mathbf{v}_s \\ &\quad \left. +\nabla\rho_s\cdot\nabla\otimes\nabla\rho_s\left(\xi_a\mathbf{v}_a\right)\right]\,dV\end{aligned} \quad (82)$$

Consider the following identities for a second order tensor fiel **A**, a third order tensor fiel $\mathfrak{A}$[6] and a vector fiel **v**

$$\begin{aligned}\mathbf{A}\cdot\nabla\mathbf{v} &= div\left(\mathbf{A}^T\mathbf{v}\right)-\mathbf{v}\cdot div\mathbf{A}, \\ \mathfrak{A}\cdot\nabla\otimes\nabla\mathbf{v} &= div\left(\mathfrak{A}^T\,\nabla\mathbf{v}\right)-\nabla\mathbf{v}\cdot div\mathfrak{A} \\ &= div\left(\mathfrak{A}^T\,\nabla\mathbf{v}\right)-div\left[\left(div\mathfrak{A}\right)^T\mathbf{v}\right]+\mathbf{v}\cdot div\left(div\mathfrak{A}\right),\end{aligned} \quad (83)$$

using these identities in (82) for $\mathbf{v}=\mathbf{v}_s$, **A** and $\mathfrak{A}$ being the coefficient of $\nabla\mathbf{v}_s$ and $\nabla\otimes\nabla\mathbf{v}_s$ in (82) respectively, the aforementioned equation takes the alternative form

$$\begin{aligned}\frac{1}{\lambda_s}I_{add} &= \int_{\mathcal{B}}\left\{div\left[\left[\left(\frac{1}{2}f_{ss}\xi_s\mathbf{I}-f_{ss}\mathbf{I}-\nabla\rho_s\otimes\nabla\rho_s\right)\mathbf{v}_s\right]+\frac{1}{2}f_{ss}\xi_f\mathbf{v}_f\right.\right. \\ &\quad \left.-\rho_s\left(\mathbf{I}\otimes\nabla\rho_s\right)\nabla\mathbf{v}_s+\left[div\left(\rho_s\nabla\rho_s\otimes\mathbf{I}\right)\right]^T\mathbf{v}_s\right] \\ &\quad -\mathbf{v}_s\cdot\left[\frac{1}{2}\xi_f\nabla f_{ss}-\frac{1}{2}f_{ss}\nabla\xi_s+div\left(\frac{1}{2}f_{ss}\xi_s\mathbf{I}-f_{ss}\mathbf{I}-\nabla\rho_s\otimes\nabla\rho_s\right)\right. \\ &\quad \left.\left. +div\,div\left(\rho_s\nabla\rho_s\otimes\mathbf{I}\right)\right]\right\}\,dV\end{aligned} \quad (84)$$

Using the divergence theorem where appropriate yields formula (45).

Combining this result with the "firs gradient " expression of (40) – using (42) and (43) yields now

$$\begin{aligned}\frac{d}{dt}\int_{\mathcal{B}}\psi^{int}\,dV &= \int_{\mathcal{B}}\left\{\left[\nabla p_s-\mathbf{m}_s-\lambda_s\,div\left(\rho_s\triangle\rho_s\mathbf{I}+\frac{1}{2}f_{ss}\mathbf{I}-\nabla\rho_s\otimes\nabla\rho_s\right)\right]\cdot\mathbf{v}_s\right. \\ &\quad \left.+\left(\nabla p_f-\mathbf{m}_f\right)\cdot\mathbf{v}_f\right\}\,dV \\ &\quad +\int_{\partial\mathcal{B}}\left\{\left[-p_s\mathbf{n}+\lambda_s\left(\rho_s\triangle\rho_s\mathbf{I}+\frac{1}{2}f_{ss}\mathbf{I}-\nabla\rho_s\otimes\nabla\rho_s\right)\mathbf{n}\right]\cdot\mathbf{v}_s\right. \\ &\quad \left.+\left(-p_f\,\mathbf{n}+\frac{\lambda_s}{2}f_{ss}\right)\cdot\mathbf{v}_f-\lambda_s\rho_s\left(\nabla\rho_s\cdot\mathbf{n}\right)\mathbf{I}\cdot\nabla\mathbf{v}_s\right\}\,dA,\end{aligned} \quad (85)$$

---

[6] We assume that a third order tensor $\mathfrak{A}$ is a linear map define as follows

$$\mathfrak{A}:\mathscr{V}\to LIN\left(\mathscr{V},LIN\left(\mathscr{V}\right)\right)$$

$\mathscr{V}$ being a linear space, $LIN\left(\mathscr{V}\right)$ the collection of all linear endomorphisms on $\mathscr{V}$ and $LIN\left(\mathscr{V},LIN\left(\mathscr{V}\right)\right)$ the collection of all the linear morphisms mapping $\mathscr{V}$ into $LIN\left(\mathscr{V}\right)$.

The transpose of $\mathfrak{A}$ is assumed to fulfil the following relation

$$\mathfrak{A}\mathbf{u}\cdot\mathbf{U}=\mathbf{u}\cdot\mathfrak{A}^T\mathbf{U}$$

for any $\mathbf{u}\in\mathscr{V}$ and any $\mathbf{U}\in LIN\left(\mathscr{V}\right)$.

from which the local statements (46) and (47) are now readily deduced.

## Appendix B

In this Appendix we derive formulas (57) using (52) and (53). To this end, one needs

$$\begin{aligned} div \mathfrak{B}_s &= -div\left(\rho_s \frac{\partial \psi^{ext}}{\partial(\nabla \rho_s)} \otimes \mathbf{I}\right) = div\left(\rho_s \frac{\partial \psi^{ext}}{\partial(\nabla \rho_s)}\right)\mathbf{I} \\ div_s(\mathfrak{B}_s \mathbf{n}) &= -div_s\left(\rho_s \frac{\partial \psi^{ext}}{\partial(\nabla \rho_s)} \cdot \mathbf{n}\,\mathbf{I}\right) = -\left(\rho_s \frac{\partial \psi^{ext}}{\partial(\nabla \rho_s)} \cdot \mathbf{n}\right) div_s \mathbf{I} \\ &\quad - \nabla^s\left(\rho_s \frac{\partial \psi^{ext}}{\partial(\nabla \rho_s)} \cdot \mathbf{n}\right) \\ &= -\left(\rho_s \frac{\partial \psi^{ext}}{\partial(\nabla \rho_s)} \cdot \mathbf{n}\right)\nabla^s \mathbf{n} - \nabla^s\left(\rho_s \frac{\partial \psi^{ext}}{\partial(\nabla \rho_s)} \cdot \mathbf{n}\right). \end{aligned} \tag{86}$$

Inserting these expressions in the formulas (53) one obtains

$$\begin{aligned} \mathbf{t}_s &= \left[-\frac{\partial \psi^{ext}}{\partial \rho_s}\rho_s + \xi_s \psi^{ext} - \frac{\partial \psi^{ext}}{\partial(\nabla \rho_s)} \cdot \nabla \rho_s + div\left(\rho_s \frac{\partial \psi^{ext}}{\partial(\nabla \rho_s)}\right) + \right. \\ &\quad \left. + \rho_s \frac{\partial \psi^{ext}}{\partial(\nabla \rho_s)} \cdot \mathbf{n}\,(tr\,\nabla^s \mathbf{n}) - \left(\frac{\partial \psi^{ext}}{\partial(\nabla \rho_s)} \cdot \mathbf{n}\right)\frac{\partial \rho_s}{\partial \mathbf{n}}\right]\mathbf{n} + \rho_s \nabla^s\left(\frac{\partial \psi^{ext}}{\partial(\nabla \rho_s)} \cdot \mathbf{n}\right), \\ \mathbf{t}_f &= \left(-\frac{\partial \psi^{ext}}{\partial \rho_f}\rho_f + \xi_f \psi^{ext}\right)\mathbf{n}, \\ \tau_s &= -\left(\rho_s \frac{\partial \psi^{ext}}{\partial(\nabla \rho_s)} \cdot \mathbf{n}\right)\mathbf{n}, \\ \tau_f &= \mathbf{0}, \end{aligned} \tag{87}$$

which agrees with (57).

*Acknowledgements.* The authors wish to thank Prof. Pierre Seppecher from Université de Toulon et du Var for his constructive criticism and long discussions about conservative boundary conditions in mixture theories. They also acknowledge the constructive reviews of two referees.

## References

1. Bowen RM (1980) Incompressible porous media models by use the theory of mixtures. Int. J. Engng. Sci. 18, 1129–1184
2. Bowen RM (1982) Compressible porous media models by use of the theory of mixtures. Int. J. Engng. Sci. 20, 697–734
3. Casal P (1961) La Capillarité interne. Cahier du groupe Francais d'Etudes de Rheologie C.N.R.S. VI(3), 31–37
4. Coddington EA, Levinson N (1955) Theory of ordinary differential equations. New York: Mc Graw-Hill
5. Cosenza P (1996) Sur les couplages entre comportement mécanique et processus de transfert de masse dans le sel gemme. Thèse Université. PARIS VI
6. Coussy O (1995) Mechanics of porous media. New York: John Wiley and Sons
7. de Boer R, Ehlers W (1990) The development of the concept of effective stress. Acta Mechanica 83, 77–92
8. de Boer R (2000) Theory of porous media. Berlin Heidelberg New York: Springer
9. dell'Isola F, Guarascio M, Hutter K (2000) A variational approach for the deformation of a saturated porous solid. A second gradient theory extending Terzaghi's effective stress principle. Archive of Applied Mechanics 70, 323–337
10. dell'Isola F, Hutter K (1998) A qualitative analysis of the dynamics of a sheared and pressurized layer of saturated soil. Proc. R. Soc. Lond.A 454, 3105–3120
11. dell'Isola F, Hutter K (1999) Variations of porosity in a sheared pressurized layer of saturated soil induced by vertical drainage of water Proc. R. Soc. Lond.A 455, 2841–2860
12. Germain P (1973) La méthode des puissances virtuelles en mécanique des milieux continus Journal de Mécanique,12(2), 235–274
13. Gouin H (1991) Variational methods for flui mixtures of grade n Continuum Models and Discrete Systems 2, 243–252
14. Hutter K, Jöhnk K, Svendsen B (1994) On interfacial transition conditions in two phase gravity flow ZAMP, 45 746–762
15. Morland LW (1972) A simple constitutive theory for a flui saturated porous solid, J. Geoph. Res. 77, 890–900
16. Müller I (1985) Thermodynamics, Pittman, Boston, etc.

17. Rajagopal KR, Tao L (1995) Mechanics of Mixtures, World Scientifi
18. Rajagopal KR, Wineman AS, Gandhi MV (1986) On boundary conditions for a certain class of problems in mixture theory,. Int. J. Engng. Sci. 24, 1453–1463
19. Seppecher P (1987) Etude d'une modelisation des zones capillaires fluides  interfaces et lignes de contact. Thèse Université. PARIS VI
20. Seppecher P (1989) Etude des condition aux limites en théorie du second gradient: cas de la capillarité C. R. Acad. Sci. Paris 309 (Séries II) 597–502
21. Svendsen B, Gray JMNT (1996) Balance relations for classical mixtures containing a moving non-material surface. Continuum Mechanics and Thermodynamics 8, 171–187
22. Svendsen B, Hutter K (1995) On the thermodynamics of a mixture of isotropic materials with constraints. Int. J. Engng. Sci. 33, 2021–2054
23. von Terzaghi K (1946) Theoretical Soil Mechanics. New York: John Wiley and Sons